\documentclass[twocolumn,aps,preprintnumbers,footinbib,floatfix,superscriptaddress]{revtex4}
\usepackage{graphicx}
\usepackage{epstopdf}
\usepackage{bm,bbold}
\usepackage{color,}
\usepackage{xcolor, soul}
\usepackage{amsfonts}
\usepackage{amsmath}
\usepackage[percent]{overpic}
\usepackage[colorlinks=true,citecolor=blue]{hyperref}

\hypersetup{colorlinks=true,citecolor=blue,linkcolor=red,urlcolor=blue}
\begin{document}
\title{Impurity effects and bandgap-closing in massive Dirac systems}
\date{\today}
\author{Habib Rostami }
\affiliation{Istituto Italiano di Tecnologia, Graphene Labs, Via Morego 30, I-16163 Genova,~Italy}
\author{Emmanuele Cappelluti}
\affiliation{Istituto dei Sistemi Complessi, CNR, 00185 Roma, Italy}
\affiliation{Dipartimento di Fisica, Universit\`a La Sapienza, P.le A. Moro 2, 00185 Roma, Italy}
\begin{abstract}
We investigate the effects in the spectral properties of a massive Dirac
system of the dynamical renormalization induce by disorder/impurity scattering
within the self-consistent Born approximation. 
We show how that these effects leads to a remarkable closing of the
bandgap edge. Above a critical value $U_c$ of the impurity scattering
the gap eventually closes, giving rise a finite density of states at
zero energy. We show that the bandgap closing stems from
the quasi-particle dynamical renormalization and
it is not associated
with the vanishing of the effective massive term.
Incoherent processes are fundamental to describe such physics.
\end{abstract}
\maketitle
\section{Introduction}
\vspace{-1.5mm}
Low dimensional electronic properties of graphene are commonly
described by the Dirac model, resulting in two Dirac cones
with linear dispersion\cite{graphene_review}. Controlling and tailoring the properties of
such dispersion, in particular in regards with the possibility
of opening a gap, is one of the most important challenges in the field,
from the theoretical as well as technological point of view.
The most straightforward way of opening a gap in graphene is by
explicitly inducing
a breaking of the sublattice
symmetry,\cite{Semenoff_1984,Novoselov_2007,Zhou_2007,Bostwick_2007,
Geim_Novoselov}
as for instance obtained in graphene/SiC,~\cite{Zhou_2007}
hexagonal Boron Nitride (h-BN),\cite{Robertson_1984,Watanabe_2014}
graphene/h-BN,\cite{Amet_2013,Woods_2014,Yankowitz_2014}
and graphene/h-BN superlattices.\cite{Hunt_2013,Yankowitz_2014,Wang_2016}
Similar physics is obtained, with a mapping pseudospin $\rightarrow$
real spin, in the surface state of 3D topological insulator
in the presence of a Zeeman field.\cite{TI_1,TI_2,TI_3}
In all these examples the breaking of symmetry 
is reflected in the appearance of a {\em massive}
term $\propto \Delta \hat{\sigma}_z$ in the Dirac model, where
the $\Delta$ represents the energy difference in the two sublattices,
mapped in the two components of the Dirac spinor. 
The same paradigmatic massive Dirac model is also ofte used to describe
single-layer transition metal dichalcogenides,\cite{Xiao_2012}
after neglecting the momentum dependence of the mass term. \cite{Rostami_2013}
Other possible single-particle sources of bandgap can come be the Haldane mass $\propto \Delta \tau_z\hat\sigma_z$ \cite{Haldane} and 
the intrinsic spin-orbit coupling $\propto \Delta s_z \tau_z\hat\sigma_z$ \cite{kane_mele} with $\tau_z=\pm$ and $s_z=\pm$ as the
valley and spin degree of freedoms, respectively.
\par
Many-body effects (e.g electron-electron and electron-phonon interaction,
impurity scattering etc.) are also crucially relevant in graphene,
giving rise to a  variety of interesting phenomena as a the minimum
conductivity at zero bias \cite{Sarma_2011}, marginal Fermi liquid~\cite{Vozmediano_1999,Sarma_2007}, the linear dependence of conductivity with
doping \cite{Sarma_2011}, plasmaronic effects in the band
dispersions~\cite{plasmaronic} and so on.
From a general point of view, however,  the presence of the (spinless) 
massless nature Dirac cones
in the graphene is protected against any single-particle and many-body perturbations,
at least as long as the interaction does not lead to a spontaneous breaking
of symmetry, usually above a finite critical
coupling strength.\cite{Appelquist_1988,Khveshchenko_1988,Khveshchenko_2004,Vafek_2008,Juricic_2009,Drut_2009,Liu_2009,Gamayun_2009,Wang_2010,Gonzalez_2010,Wang_2011,Kotov_2012,Gonzalez_20121,Gonzalez_20122,Wang_2012,Popovici_2013,Cappelluti_2014}
\par\vspace{-4mm}
Among the possible sources of interaction, disorder/impurity scattering
represents the most direct and simple case, provided a conserving
approach is enforced. Impurity/disorder effects have been hence
thoughtfully analyzed in graphene and in the massless Dirac model.~\cite{Peres_2006,Pereira_2006,Pereira_2008,Hu_2008,Dora_2008}
Relatively less attention has been devoted to the study of disorder effects
in the {\em massive} Dirac model.
However, in spite of the simplicity, the analysis of impurity effects
in the massive Dirac mode gives rise to interesting novel features.
For instance, the possibility of a bandgap closing above a threshold of
impurity scattering, was briefly mentioned in
Ref.~[\onlinecite{ando_gap}], in the simplest
paradigmatic context of a
self-consistent Born approximation (SCBA),
but a careful characterization of such bandgap closing
and of its origin has not been provided. 
Note that such scenario (observed in the simple SCBA scheme)
is qualitatively different from the onset of midgap states
in highly disordered massive Dirac models,\cite{Balatsky_2006,
Wehling_2014,Gonzalez_2012,Castro_2015} which  needs
a $t$-matrix approach to be revealed and that can be more correctly interpreted
as {\em gap filling} rather than {\em gap closing}.
\vspace{-6mm}
\par
In this paper we present a detailed analysis
of the impurity-induced bandgap closing in massive Dirac model.
Following Ref. [\onlinecite{ando_gap}],
we employ a self-consistent Born approximation
that represents the simplest minimal model to take into account
impurity/disorder scattering at the homogeneous level.
We find that such bandgap closing is not associated
with a vanishing of the effective massive term, rather to
a diverging of the quasi-particle dynamical renormalization.
The role of inchoerent processes is also discussed, and
the consistency of the present results
with Luttinger's theorem validated.
\vspace{-5mm}
\par
The paper is organized in three sections. In Section~\ref{sec:model} we give the general formalism of SCBA in Dirac model. 
In Section~\ref{sec:gap} we discuss the effects of
the dynamical bandgap renormalization and of the quasiparticle
properties.
Finally, in Section~\ref{sec:conclusion} the present results are
discussed in relation of possible theoretical scenarios.
\section{The Model}\label{sec:model}
In this paper we consider the massive Dirac model
described by the Hamiltonian: 
\begin{align}
\hat{\cal H}_{\bm k}
&=
\hbar v
\left[
 k_x\hat{\sigma}_x+k_y\hat{\sigma}_y
\right]
+
\Delta \hat{\sigma}_z,
\label{ham}
\end{align}
where the term $\propto \hat{\sigma}_z$ induced a gap $\sim 2\Delta$
at the Dirac point ${\bm k}=0$.
Mimicking the specific case of graphene, we introduce an upper momentum
cut-off $k_c$ chosen to preserve the total available phase space.
Namely, considering also the valley degeneracy $N_v=2$ in graphene,
we set $N_v \pi k_c^2=V_{\rm BZ}$, where
$V_{\rm BZ}=8\pi^2/3\sqrt{3}a^2$ is the graphene Brillouin zone
and $a$ the nearest-neighbors carbon-carbon distance.
Using a typical value $a=1.42$ \AA, we get $k_c=1.09$ \AA$^{-1}$,
which can be associated with an ultraviolet energy cut-off
$W=\hbar v k_c=7.2$ eV, so that $\hbar v |{\bm k}| \le W$.

We consider scattering on local impurity centers
with density $n_{\rm imp}$ and potential
\begin{align}
V_{\rm imp}({\bm r})
&=
\sum_i
V_i
\delta({\bm r}-{\bm R}_i),
\end{align}
where ${\bm R}_i$ are the coordinates of the lattice sites.
We assume standard Born impurity correlations
\begin{align}
&\langle V_{\rm imp}({\bm r}) \rangle
=
0,
\\
\langle
&V_{\rm imp}({\bm r})V_{\rm imp}({\bm r'})
\rangle
=
n_{\rm imp} V_{\rm imp}^2\delta({\bm r}-{\bm r}'),
\end{align}
where the average $\langle \dots \rangle$ is meant over all the space
${\bm r}$ (${\bm r'}$) and over all the impurity configurations.
For simplicity we define the parameter $\gamma_{\rm imp}=n_{\rm imp} V_{\rm  imp}^2$.
With these notations we can write the Born impurity self-energy
in the Matsubara space:
\begin{align}\label{eq:sigma}
\hat{\Sigma}(i\omega_n)
&=
\gamma_{\rm imp} N_v
\sum_{\bm k}
\hat{G}({\bm k},i\omega_n),
\end{align}
where the self-consistent Green's function reads:
\begin{align}\label{eq:green}
\hat{G}({\bm k},i\omega_n)
&=
\frac{1}{i\omega_n\hat{I}-\hat{\cal H}_{\bm k}-\hat{\Sigma}(i\omega_n)}.
\end{align}
We can expand the self-energy in Pauli matrices,
$\hat{\Sigma}=\Sigma_I \hat{I}
+\Sigma_x\hat{\sigma}_x+\Sigma_y\hat{\sigma}_y+\Sigma_z\hat{\sigma}_z$.
It is easy to see that
the self-energy is momentum-independent and that
only the terms $\Sigma_I (i\omega_n)$ and $\Sigma_z (i\omega_n)$
are not zero. The self-energy terms $\Sigma_I (i\omega_n)$, $\Sigma_z (i\omega_n)$
provide a {\em dynamical} renormalization of the one-particle
excitations, 
$i\widetilde{\omega}_n=i\omega_n-\Sigma_I(i\omega_n)$,
and of the gap function, $\Delta(i\omega_n)=\Delta+\Sigma_z(i\omega_n)$.
It is useful to introduce also the quasi-particle renormalization function
$Z(i\omega_n)=1-\Sigma_I(i\omega_n)/i\omega_n$.
\par
\begin{figure}[t]
\includegraphics[width=0.9\linewidth]{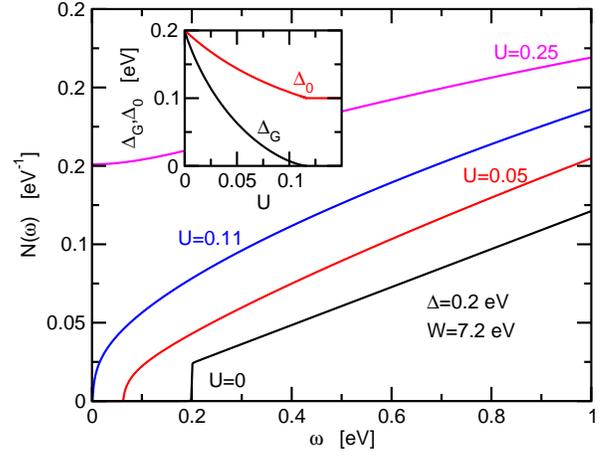}
\caption{color online) Density of states $N(\omega)$ for different
 values of the impurity scattering. We set $\Delta=0.2$ eV and
$W=7.2$ eV. The gap vanishes in the density of states
for $U$ above the critical value $U_c\approx 0.117$. Inset:
dependence of $U$ of the bandgap edge $\Delta_{\rm G}$
and of the zero-frequency gap function $\Delta_0=\Delta(\omega =0)$.}
\label{f-dos}
\end{figure}
After few straightforward steps (in Appendix \ref{a:self}),
we can write the self-consistent relations:
\begin{align}
\Sigma_I(i\omega_n)
&=
-i\omega_nZ(i\omega_n)  U\alpha(i\omega_n),
\label{e:0}
\\
\Sigma_z(i\omega_n)
&=
-\Delta(i\omega_n) U\alpha(i\omega_n),
\label{e:z}
\end{align}
where $U$ is the dimensionless parameter,
$U=N_v\gamma_{\rm imp}S_{\rm cell}/2\pi \hbar^2v^2$,
and where
\begin{align}
\alpha(i\omega_n)
&=
\ln\left[
1+\frac{W^2}{\Delta^2(i\omega_n)+\widetilde{\omega}_n^2}
\right].
\label{e:an}
\end{align}
\par
Eqs.~(\ref{e:0})-(\ref{e:an}) can be easily generalized on the
real-frequency axis, $i\omega_n \rightarrow \omega+i\eta$,
with $\eta \rightarrow 0$.
We get:
\begin{align}\label{e:zw}
\Sigma_I(\omega)
&=
-\omega Z(\omega)U\alpha(\omega),
\\
\Sigma_z(\omega)
&=
-\Delta(\omega)U\alpha(\omega),
\label{e:0w}
\end{align}
where $\Delta(\omega)=\Delta+\Sigma_z(\omega)$,
$Z(\omega)=1-\Sigma_I(\omega)/\omega$.
and $\widetilde{\omega}=\omega+i\eta-\Sigma_I(\omega)$.
All the quantities depending on the real-axis frequency
represent here the retarded part, e.g.
$\alpha(\omega)=\alpha(\omega+i\eta)$.
We have thus explicitly:
\begin{align}\label{eq:alpha}
\alpha(\omega)
&=
\ln\left[
1+\frac{W^2}{\Delta^2(\omega)-\omega^2Z^2(\omega)}
\right].
\end{align}
\par
Once known the self-energy, we can express the spectral function
$A({\bm k},\omega)$ as:
\begin{align}
A({\bm k},\omega)
&=
-\frac{1}{\pi}
\mbox{Im}\left\{
\mbox{Tr}
\left[
\hat{G}({\bm k},\omega+i\eta)
\right]
\right\}
\nonumber\\
&=
-\frac{1}{\pi} \mbox{Im} \left \{ 
\frac{2\omega Z(\omega)}
{\omega^2Z^2(\omega)-\epsilon_{\bm k}^2-\Delta^2(\omega)}
\right \} ,
\label{spectral}
\end{align}
and the density of states per spin:
\begin{align}
N(\omega)
=
N_v
\sum_{\bm k}
A({\bm k},\omega)
=
\frac{2}{W^2}
\mbox{Im}
\left[
\omega Z(\omega)\alpha(\omega)
\right].
\label{e:dos}
\end{align}
\section{Bandgap renormalization}\label{sec:gap}
Eqs.~(\ref{e:0w})-(\ref{e:zw}) have been introduced in
Ref.~[\onlinecite{ando_gap}], where also a bandgap closing was predicted,
for given $U$ impurity scattering, for values of the bare gap $\Delta$
larger than a critical value $\Delta_c(U,W)$.
Analitical estimates of $\Delta_c(U,W)$ were provided in
the asymptotic limit $U \rightarrow 0$, but the physics
underlying such bandgap closing was not  specifically addressed.

The most direct way to trace the evolution of the effective gap as a function of the impurity scattering is
the investigation of the renormalized density of state. We {\em define} thus the bandgap edge $\Delta_{\rm G}$ as
the gap observed in the DOS. Some representative cases are shown in Fig.~\ref{f-dos},
showing how the $\Delta_{\rm G}$ decreases by increasing $U$
until it vanishes above a critical value $U_c$.
The evolution of $\Delta_{\rm G}$ as a function of $U$ can be investigated in a semi-analytical way, as presented in Appendix \ref{app:bandgap_edge},
and it is depicted in the inset of Fig.~\ref{f-dos}.
A perturbation analysis 
can be also derived, predicting in the dilute impurity $U \ll U_c$ the behavior (see Appendix \ref{app:bandgap_edge})
\begin{align}
\Delta_{\rm G}
&\approx
\Delta
\left \{ 
1-2U {\cal W}\left ( \frac{ \mathbb{e}^4 W^2}{4 U \Delta^2} \right )
\right \}.
\end{align}
where $\mathbb{e}$ is the Napier's constant and  ${\cal W}(z)$ is the  Lambert ${\cal W}$-function 
(or so called ProductLog function) which obeys $ {\cal W}(z) \mathbb{e}^{{\cal W}(z)}=z$ and its asymptotic behavior is   ${\cal W}(z\gg 1)\approx \ln(z)-\ln(\ln(z))$.  
For the set of parameters used in Fig.~\ref{f-dos},
 this relation results to be accurate only for very low values of $U$ ($U<0.17U_{c}\sim0.02$), pointing out thus the compelling need of a non-perturbative
self-consistent analysis.
\par
As discussed in Ref.~[\onlinecite{ando_gap}], the critical value $U_c$ above which $\Delta_{\rm G}=0$ can be determined by the onset of an imaginary part
in the self-energy $\Sigma_I$ at $\omega=0$, $\Sigma_I(\omega=0)=-i\Gamma$ whereas the off-diagonal self-energy,
associated with the renormalization of the gap function, $\sigma=-\Sigma_z(\omega=0)$, remains a purely real quantity.
We generalize the analysis of Ref.~[\onlinecite{ando_gap}] at any value of $U$, relaxing the condition $U\ll 1$. The closing of the gap
in the density of states is thus ruled by the set of self-consistent equations for $\Gamma$ and $\sigma$:
\begin{align}
\Gamma
&=
\Gamma
U
\ln\left[
1+
\frac{W^2}
{\left(\Delta-\sigma\right)^2
+\Gamma^2}
\right],
\label{s0fw0-2}
\\
\sigma
&=
\left(\Delta-\sigma\right)
U
\ln\left[
1+
\frac{W^2}
{\left(\Delta-\sigma\right)^2
+\Gamma^2}
\right].
\label{szfw0-2}
\end{align}
Moreover the density of states at $\omega=0$ is given by
\begin{align}
N(0)
&=
\frac{2\Gamma}{W^2}
\ln\left[
1+\frac{W^2}{(\Delta-\sigma)^2+\Gamma^2}
\right].
\end{align}
\par
Eqs.~(\ref{s0fw0-2})-(\ref{szfw0-2}) admit thus two possible
solutions, for $U\le U_c$ and for $U \ge U_c$, where 
\begin{align}
U_c
&=
\frac{1}
{
\ln\left[
1+\frac{\displaystyle 4W^2}{\displaystyle \Delta^2}
\right]
}.
\label{uc}
\end{align}
\begin{figure}[t]
\includegraphics[width=0.9\linewidth]{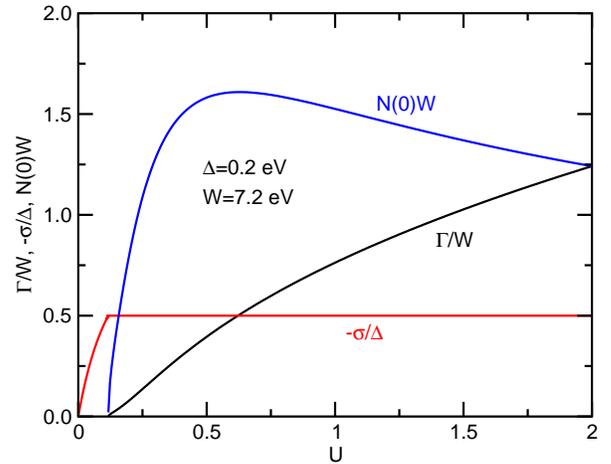}
\caption{color online) Dependence on the impurity scattering
parameter $U$ of the zero-frequency quantities $\Gamma$, $\sigma$
and $N(0)$. We set $\Delta=0.2$ eV and
$W=7.2$ eV. For these values we have $U_c=0.117$.}
\label{f-GsN_vs_U}
\end{figure}
For weak impurity scattering $U \le U_c$ we have thus $\Gamma=0$ and finite $\sigma \le \Delta/2$, implying zero density of states $N(0)=0$,
whereas, for $U \ge U_c$, $\sigma$ gets locked at $\sigma=\Delta/2$ and $\Gamma$ finite, with $\Gamma \propto \sqrt{U-U_c}$
for $U \rightarrow U_c^+$. An analytical expression for the density of states at zero energy can be obtained in this regime:
\begin{align}
N(0)
&=
\frac{2\Gamma(U,\Delta,W)}{UW^2},
\end{align}
where
\begin{align}
\Gamma
=
W \Theta(U-U_c)
\sqrt{
\frac{1}{\exp(1/U)-1}
-
\frac{1}{\exp(1/U_c)-1}
}.
\end{align}
where $\Theta(x)$ is the Heaviside function.  The plot of $\Gamma$, $\sigma$ and $N(0)$
as functions of $U$ is shown in Fig.~\ref{f-GsN_vs_U}. We also obtain the asymptotic dependences
$\Gamma$, $N(0) \propto \sqrt{U-U_c}$, for $U \rightarrow U_c^+$, and $\Gamma \propto \sqrt{U}$,
$N(0) \propto 1/\sqrt{U}$, for $U \gg U_c$. 
\par
The value $\Delta_0=\Delta(\omega=0)=\Delta-\sigma$
represents the value of the gap function $\propto \hat{\sigma}_z$
at zero energy excitations, in the presence of disorder/impurity scattering.
We note that, as shown in the inset of Fig.~\ref{f-dos},
unlike $\Delta_{\rm G}$, $\Delta_0$ is always finite for {\em any} value of $U$,
signalizing that, in the spirit of a renormalization group approach,
a finite term $\propto \hat{\sigma}_z$ is always relevant in the
low-energy limit $\omega \rightarrow 0$, in agreement
with the topological robustness of the massive Dirac model
in the presence of a scattering source that does not affect the symmetry
properties.
\par
This observation points out that the closing of the bandgap $\Delta_{\rm G}$ cannot be understood in terms of purely {\em topological} excitations $\propto \hat{\sigma}_z$
associated with $\Delta(\omega)$, but the renormalization of the quasi-particle spectral properties $\propto \hat{\sigma}_I$, associated with $Z(\omega)$, plays a fundamental role.
\begin{figure}[t]
\includegraphics[scale=0.53,clip=]{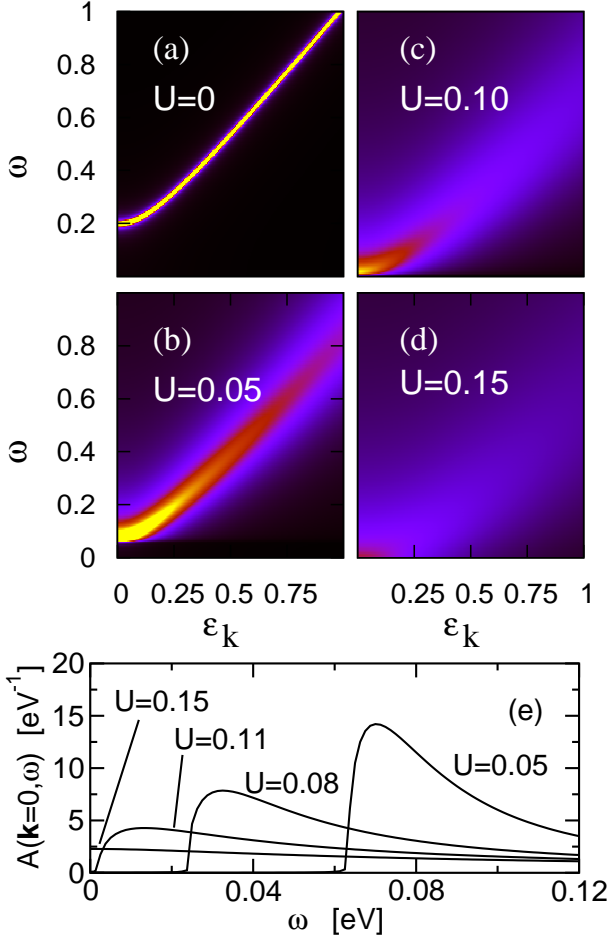}
\includegraphics[scale=0.43,clip=]{fig3_bottom}
\caption{(color online) (a)-(d) Intensity map of the spectral function
$A({\bm k},\omega)$ for ($U < U_c$) $U=0$, $U=0.05$, $U=0.10$ and ($U > U_c$) $U=0.15$.
Here $\Delta=0.2$ eV, $W=7.2$ eV and $\epsilon_{\bm k}=\hbar v |{\bm k}|$.
(e) Spectral function $A({\bm k},\omega)$ at $\epsilon_{\bm k}=0$ for $U=0.05, 0.08, 0.11, 0.15$.}
\label{f-map_g0.2_u}
\end{figure}
\par
In order to shed further light on this issue, we analyze the properties of the spectral function, defined in Eq.~(\ref{spectral}), which can be conveniently rewritten as:
\begin{align}
A({\bm k},\omega)
=
-\frac{2}{\pi}
\mbox{Im}\left \{
\frac{\omega/Z(\omega)}
{\omega^2-[\epsilon_{\bm k}/Z(\omega)]^2-\widetilde{\Delta}(\omega)^2}
\right\} ,
\label{spectral2}
\end{align}
where $\widetilde{\Delta}(\omega)=\Delta(\omega)/Z(\omega)$.
The function $Z(\omega)$ is associated with the renormalization of the
one-particle spectral weight and of the one-particle dispersion,
whereas $\widetilde{\Delta}(\omega)$ rules the effective excitations in
the presence of the gap {\em and} of the dynamical renormalization induced by the
scattering processes.
\begin{figure}[t]
\includegraphics[scale=0.43,clip=]{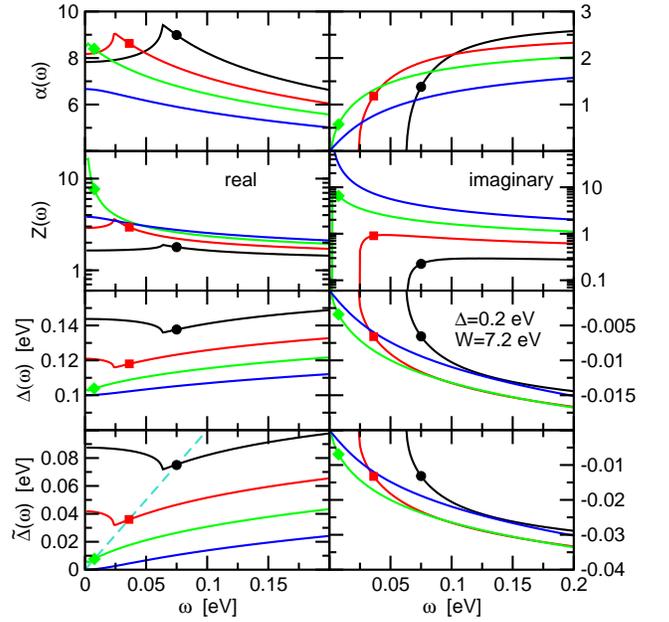}
\caption{(color online) Real parts (left panels) and imaginary parts (right panels) of the complex functions $\alpha(\omega)$.,
$Z(\omega)$, $\Delta(\omega)$, $\widetilde{\Delta}(\omega)$, for few representative $U$: $U=0.05$ (black), $U=0.08$ (red), 
$U=0.11$ (green), $U=0.15$ (blue). A coherent gap is determined, for $U \le U_c$, in the left-bottom panel by the graphical solution of Eq.~(\ref{gapcoh}),
for $U=0.05$ (black circle), $U=0.08$ (red square), and $U=0.11$ (green diamond). No coherent gap is defined for $U=0.15 > U_c$.
The value of the quantities (real and imaginary parts) for $\omega=\widetilde{\Delta}_{\rm coh}$ in the other panels is also shown by the corresponding symbols.}
\label{f-g0.2_u}
\end{figure}
\begin{figure}[t]
\includegraphics[scale=0.43,clip=!]{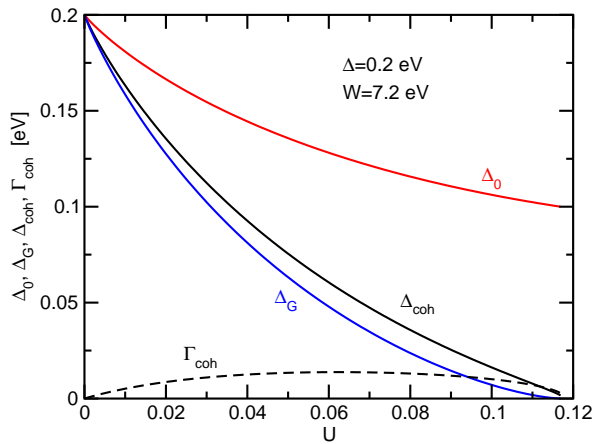}
\caption{(color online) Plot of ``coherent'' bandgap $\widetilde{\Delta}_{\rm coh}$, of the true bandgap $\Delta_{\rm G},$ of the zero-frequency gap function $\Delta_0$
as function of the dimensionless impurity scattering parameter $U$. Also shown is the self-consistent damping $\Gamma_{\rm coh}$ of the coherent bandgap, as defined in Eq.~(\ref{gapimg})~.}
\label{f-gaps}
\end{figure}
\par
The intensity map of the spectral function $A({\bm k},\omega)$ for few representative cases in the regimes $U < U_c$ and $U > U_c$ is shown in Fig.~\ref{f-map_g0.2_u}(a)-(d).
A dispersive ``quasi-particle'' peak in the spectral intensity is detected at low energy  for all $U \le
U_c$, whereas any signature of coherent peak disappears for $U \ge U_c$. Note that the ``quasi-particle'' peak follows the dependence of
$\Delta_{\rm G}$ rather than $\Delta_0$.
A similar behavior is traceable in the evolution of the spectral function at the bandgap edge $\epsilon_{\bm k}=0$, as shown
in Fig.~\ref{f-map_g0.2_u}(e). Most noticeable is here the anomalous (non-Lorentzian) shape of the spectra function, even for $U \le U_c$ signalizing that 
the coherent quasi-particle picture is questionable and incoherent processes associated with the imaginary parts of the self-energy are here relevant.
\par
To investigate in more detail this issue we plot in Fig.~\ref{f-g0.2_u} the real (left panels) and imaginary (right panels) parts of the relevant
quantities $Z(\omega)$,  $\Delta(\omega)$,
$\widetilde{\Delta}(\omega)$, as well as the function
$\alpha(\omega)$.
\par
In the spirit of a picture of quasi-particle coherent excitations, we can {\em define} a coherent bandgap $\widetilde{\Delta}_{\rm coh}$ at $\epsilon_{\bm k}=0$
by {\em assuming} the imaginary part of the denominator of Eq.~(\ref{spectral2}) to be negligible, and by looking for the zeroes of the real part, i.e.
\begin{align}
\omega
=\mbox{Re}[\widetilde{\Delta}(\omega)]
\Big |_{\omega=\widetilde{\Delta}_{\rm coh}}.
\label{gapcoh}
\end{align}
The value of the coherent bandgap $\widetilde{\Delta}_{\rm coh}$ so obtained, for some relevant values of the impurity scattering, is marked by filled symbols in the left-lower panel,
and the values of all the other relevant functions at such energy
$\omega=\widetilde{\Delta}_{\rm coh}$ is shown in the other panels.
\par
The resulting dependence of $\widetilde{\Delta}_{\rm coh}$ as a function of $U$, compared with $\Delta_{\rm G}$ and $\Delta_0$,
is shown in Fig.~\ref{f-gaps}.The fair agreement between $\widetilde{\Delta}_{\rm coh}$ and $\Delta_{\rm G}$ shows that, except close to $U \approx U_c$,
the gap states can be at a good extent described as quasi-coherent excitations, in accordance with the presence of bright quasi-particle peaks for 
$U \le U_c$ in Fig.~\ref{f-map_g0.2_u}. We can further check the robustness of the coherent-gap approximation, as defined in Eq.~(\ref{gapcoh}), 
by evaluating the expected imaginary part of the gap-function,
\begin{align}
\Gamma_{\rm coh}
=
\left.
\mbox{Im}[\widetilde{\Delta}(\omega)]
\right|_{\omega=\widetilde{\Delta}_{\rm coh}}
.
\label{gapimg}
\end{align}
\par
As expected, $\Gamma_{\rm coh} \ll \widetilde{\Delta}_{\rm  coh}$
for any $U \le U_c$ (except very close to $U_c$),
supporting the validity of the quasi-coherent gap approximation.
On the other hand, although relatively small, a finite imaginary part
$\Gamma_{\rm coh}$ is also always associated with such excitations,
giving rise to the asymmetric shape of the spectral function.
\par
Eventually, however, for $U \approx U_c$, the incoherent contribution
become dominant, $\Gamma_{\rm coh} \gg \widetilde{\Delta}_{\rm  coh}$,
and the effective bandgap closing cannot be described without such
incoherent processes.
\section{Discussion and conclusions}\label{sec:conclusion}
In this paper we have investigated, within a self-consistent Born
approximation scheme, the effects of disorder/impurity scattering
on the one-particle spectral properties of gapped Dirac model.
We have shown that the dynamical self-energy 
leads to a renormalization (reduction) of the bandgap edge.
For impurity scattering above a critical value, $U \ge U_c$
the bandgap closes and a finite density of states appears
at zero energy.
We have shown that such bandgap closing does not stem from
a vanishing of the massive term, but it is rather associated
with a divergence of the one-particle dynamical renormalization
function.
Although a quasi-particle picture can be a good approximation
in a wide range of the impurity scattering $U$,
we have shown a finite incoherent weight
is always present.
\par
The onset of a finite density of states at zero energy,
associated with the closing of the bandgao edge,
above a threshold $U\ge U_{\rm c}$,
might appear at odds with the Luttinger's theorem,
according which, in the absence of a phase transition,
the area of the Fermi surface
(and hence the DOS, for ${\bm  k}$-independent self-energies)
should be conserved for any value of the scattering interaction.
It is however important to 
underline here that, as we have remarked in the previous Section, 
a finite imaginary part of the self-energy is strongly tight with
the presence of a finite density of states, so that
no perfectly coherent excitation with zero damping can be defined,
even with zero energy $\omega=0$.
From another point of view, this implies that, technically speaking,
a Fermi liquid cannot be defined, questioning thus one of the basic
assumptions for the validity of the Luttinger's theorem. \cite{Luttinger_Ward,Luttinger}
The onset of a finite density of states for $U\ge U_{\rm c}$,
is thus not in contradiction with the Luttinger's theorem,
that cannot be applied here.
\appendix
\section{Impurity self-energy}\label{a:self} 
We summarize in the present Appendix the details of the derivation
of the impurity self-energy in the Born approximation.
By using Eqs.~(\ref{eq:sigma}) and (\ref{eq:green}),
we can immediately write
the self-consistent relations for the impurity self-energy
in the imaginary-frequency Matsubara space:
\begin{align}
\Sigma_I(i\omega_n)
&=
-
i\widetilde{\omega}_n U
\int_0^W 
\frac{2\epsilon d\epsilon}
{\widetilde{\omega}_n^2+\epsilon^2+ {\Delta}(i\omega_n)^2},
\label{ea:0}
\\
\Sigma_z(i\omega_n)
&=
-
 {\Delta}(i\omega_n) U
\int_0^W 
\frac{2\epsilon d\epsilon}
{\widetilde{\omega}_n^2+\epsilon^2+ {\Delta}(i\omega_n)^2},
\label{ea:z}
\end{align}
where $i\widetilde{\omega}_n=i\omega_n-\Sigma_I(i\omega_n)$,
$ {\Delta}(i\omega_n)=\Delta+\Sigma_z(i\omega_n)$,
and where we have introduced a ultraviolet cut-off $W$
for the linear Dirac dispersion $\epsilon=\hbar v k$.
Writing $\sum_{\bm k} =(S_{\rm cell}/2\pi \hbar^2v^2)\int d\epsilon
\epsilon \int d\theta/2\pi$,
we have also defined here the dimensionless parameter
$U=N_v\gamma_{\rm imp}S_{\rm cell}/2\pi \hbar^2v^2$.
Eqs. (\ref{ea:0})-(\ref{ea:z}) can be also written in the more compact form:
\begin{align}
\Sigma_I(i\omega_n)
&=
-
i\widetilde{\omega}_n U
\alpha(i\omega_n)
\nonumber\\
&=
-i\omega_n\frac{U\alpha(i\omega_n)}
{1-U\alpha(i\omega_n)},
\label{ea:02}
\\
\Sigma_z(i\omega_n)
&=
-
 {\Delta}(i\omega_n) U
\alpha(i\omega_n)
\nonumber\\
&=
-i\Delta\frac{U\alpha(i\omega_n)}
{1+U\alpha(i\omega_n)},
\label{ea:z2}
\end{align}
where
\begin{align}
\alpha(i\omega_n)
&=
\int_0^W 
\frac{2\epsilon d\epsilon}
{\widetilde{\omega}_n^2+\epsilon^2+ {\Delta}(i\omega_n)^2}.
\end{align}

\section{Bandgap edge $\Delta_{\rm G}$  and bandgap closing at $U_{\rm c}$}\label{app:bandgap_edge}
In this Appendix we provide some semi-analytical expressions for the
evolution
of the bandgap and of the density of states at zero energy $N(0)$
as functions of the impurity scattering.

For a generic impurity scattering parameter $U$,
the density of states is given by Eq. (\ref{e:dos})
once the complex quantities $Z(\omega)$, $\Delta(\omega)$,
$\alpha(\omega)$ are obtained by the self-consistent solution
of Eqs. (\ref{e:0w}-\ref{eq:alpha}).
A finite density of states $N(\omega)$ at $|\omega| \le  \Delta_{\rm G}$
is associated with a finite imaginary part of the functions
$Z(\omega)$, $\Delta(\omega)$,
$\alpha(\omega)$, whereas the gapped region $|\omega| \ge \Delta_{\rm  G}$
is characterized by the existence of a solution of
Eqs. (\ref{e:0w}-\ref{eq:alpha})
with purely real quantities $Z(\omega)$, $\Delta(\omega)$,
$\alpha(\omega)$.

To evaluate $\Delta_{\rm  G}$, let us assume thus, for given $\omega$,
purely real quantities $Z(\omega)=Z_{\rm R}$, $\Delta(\omega)=\Delta_{\rm R}$,
$\alpha(\omega)=\alpha_{\rm R}$.
They should fulfill the set of equations:
\begin{align}
\alpha_{\rm R}
&=
\ln\left[
1+\frac{W^2}{\Delta_{\rm R}^2-\omega^2Z_{\rm R}^2}
\right],
\label{e:ap2}
\\
\Delta_{\rm R}
&=
\frac{\Delta}{1+U\alpha_{\rm R}},
\label{d:ap2}
\\
Z_{\rm R}
&=
\frac{1}{1-U\alpha_{\rm R}}.
\label{z:ap2}
\end{align}

Eqs. (\ref{e:ap2})-(\ref{z:ap2})
can be conveniently recast
in a more compact form as:
\begin{align}
A\left(\alpha_{\rm R}\right)
&=
B\left(\alpha_{\rm R}\right),
\label{my1}
\end{align}
where
\begin{align}
A\left(\alpha_{\rm R}\right)
&=
\exp\left(\alpha_{\rm R}\right),
\label{my2}
\\
B\left(\alpha_{\rm R}\right)
&=
1+\frac{W^2(1+U\alpha_{\rm R})^2}{\Delta^2
-\widetilde {\omega}_{\rm R}^2(1+U\alpha_{\rm R})^2},
\label{my3}
\end{align}
and where
\begin{align}
\widetilde {\omega}_{\rm R}
&=
\omega Z_{\rm R}.
\label{my4}
\end{align}

\begin{figure}[t]
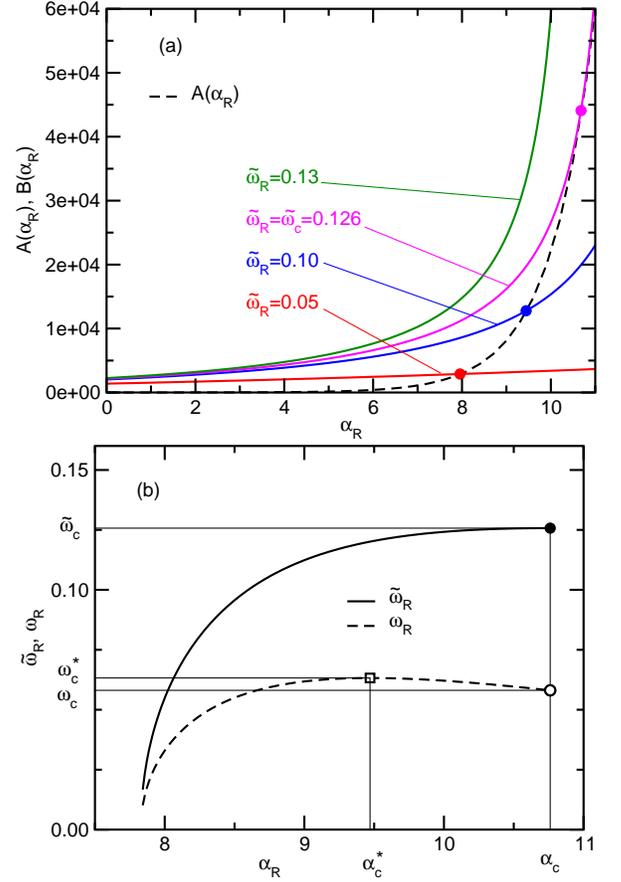

\includegraphics[scale=0.43,clip=]{f-a.eps}
\includegraphics[scale=0.43,clip=]{f-ac.eps}
\caption{(color online) (a) Graphical solution of Eq. (\ref{my1}) for
$\Delta=0.2$ eV and $U=0.05$.
The black dashed line represents the left hand side of Eq. (\ref{my1}),
$A\left(\alpha_{\rm R}\right)$, while the colored solid lines the
right hand side $B\left(\alpha_{\rm R}\right)$ for different values
of $\widetilde {\omega}_{\rm R}$. A real solution (marked by filled
circles) exists for 
$\widetilde {\omega}_{\rm R} \le \widetilde {\omega}_{\rm c}$,
corresponding to a gapped states.
(b) Plot of $\widetilde {\omega}_{\rm R}$ and $\omega_{\rm R}$
vs. $\alpha_{\rm R}$. The values $\alpha_c$, 
$\widetilde {\omega}_c$ and $\omega_c$ are marked by empty and filled
circles.
Note that, due to the non-monotonic behavior,
the maximum extension of $\omega_{\rm R}$,
defining the effective bandgap $\Delta_{\rm G}$,
is associated with a value $\omega_c^* > \omega_c$.
The value of $\omega_c^*$. is marked by a empty square.
}
\label{f-fa}
\end{figure}

Eq. (\ref{my1}), supported with 
Eqs. (\ref{my2})-(\ref{my4}), can be 
solved now in a self-consistent way for given $\widetilde
{\omega}_{\rm R}$.
The graphical solution for different values of $\widetilde{\omega}_{\rm R}$
is shown in Fig. \ref{f-fa}a.
Real solutions (marked by filled circles) exist for 
$\widetilde {\omega}_{\rm R} \le \widetilde {\omega}_{\rm c}$,
signalizing for these frequencies zero density of states and hence
a bandgap, whereas
a finite density of states appears for
$\widetilde {\omega}_{\rm R} \ge \widetilde {\omega}_{\rm c}$,
were no possible real solution of Eqs. (\ref{my1})-(\ref{my3})
exists and a finite imaginary part/density of states is enforced.
As clear from Fig. \ref{f-fa}a, the critical value ${\omega}_{\rm c}$,
can be determined by the condition
\begin{align}
A'\left(\alpha_{\rm R}\right)
&=
B'\left(\alpha_{\rm R}\right),
\end{align}
which, together with Eqs. (\ref{my1})-(\ref{my4}),
after few straightforward mathematical steps, provides
an analytical implicit expression
\begin{align}
\widetilde {\omega}_{\rm c}
&=
\sqrt{\Delta_{\rm c}^2-\frac{W^2}{\exp(\alpha_{\rm c})-1}},
\end{align}
where there is a self-consistent relation between $\Delta_{\rm c}$
and $\alpha_{\rm c}$: 
\begin{align}
1
&=
8U\frac{\Delta_{\rm c}^3}{\Delta W^2}
\sinh^2 \left(\alpha_{\rm c}/2\right),
\\
\Delta_{\rm c}
&=
\frac{\Delta}{1+U\alpha_{\rm c}}.
\end{align}

Finally, one can write:
\begin{align}
\omega_{\rm c}
&=
\widetilde {\omega}_{\rm c}
\left[1-U\alpha_{\rm c}\right].
\end{align}

The plot of $\omega_{\rm R}$, $\widetilde {\omega}_{\rm R}$
as functions of $\alpha_{\rm R}$ for the representative case
$\Delta=0.2$ eV, $U=0.05$ is shown in Fig. \ref{f-fa}b.
The filled circle marks here the values
of $\widetilde {\omega}_{\rm c}$  and $\alpha_{\rm c}$,
determining as well to the value of $\omega_{\rm c}$
(empty circle).
A important feature to be underlined here, however, is that
there is a monotonic one-to-one dependence between
$\widetilde {\omega}_{\rm R}$ and $\alpha_{\rm R}$,
but no monotonic dependence of $\omega_{\rm R}$
as a function of $\alpha_{\rm R}$.
The real bandgap edge $\Delta_{\rm G}$ can be thus identified
with the largest obtainable value of $\omega_{\rm R}$ which,
as evident from Fig. \ref{f-fa}b, is not given
by $\omega_{\rm c}$, but rather by 
$\Delta_{\rm G}=\omega_{\rm c}^*$, where $\omega_{\rm c}^*$
can be obtained by the condition $d\omega_{\rm R}/d\alpha_{\rm R}=0$.
This condition can be also seen in Fig.~\ref{f-g0.2_u} where the slope  of $\alpha(\omega)$ function diverges when $\omega$ approach $\Delta_{\rm G}$ from below.

Such analytical constraint can be enforced by
using Eqs. (\ref{my1})-(\ref{my4}).
After few technical steps, we can write
the implicit relation
\begin{align}
0
&=
-
2U\left[1-U\alpha_{\rm c}^*\right]
\left[
\Delta_*^2-\frac{W^2}{\exp(\alpha_{\rm c}^*)-1}
\right]
\nonumber\\
&
-
2U\left[1-U\alpha_{\rm c}^*\right]^2\frac{\Delta_*^3}{\Delta}
\nonumber\\
&
+
\left[1-U\alpha_{\rm c}^*\right]^2W^2
\frac{\exp(\alpha_{\rm c}^*)}{[\exp(\alpha_{\rm c}^*)-1]^2},
\label{newapp1}
\end{align}
where
\begin{align}
\Delta_*
&=
\frac{\Delta}{1+U\alpha_{\rm c}^*}.
\label{newapp2}
\end{align}
In addition, we have
\begin{align}
\widetilde {\omega}_{\rm c}^*
&=
\sqrt{\Delta_*^2-\frac{W^2}{\exp(\alpha_{\rm c}^*)-1}},
\label{newapp3}
\end{align}
and
\begin{align}
\omega_{\rm c}^*
&=
\widetilde {\omega}_{\rm c}^*
\left[1-U\alpha_{\rm c}^*\right].
\label{newapp4}
\end{align}
The value $\omega_{\rm c}^*$ in Eq. (\ref{newapp2}),
along with the self-consistent solution
of Eqs. (\ref{newapp1}), (\ref{newapp3}), (\ref{newapp4}),
identifies the bandgap $\Delta_{\rm G}(U)=\omega_{\rm c}^*$
as a function of the impurity scattering parameter $U$.

A perturbation approach can be also applied to extract analytically the
magnitude of the bandgap reduction in the
dilute impurity $U \ll U_c$ regime.
After few straightforward steps, we obtain thus:
\begin{align}
\Delta_{\rm G}
&\approx
\Delta
\left \{ 
1-2U {\cal W}\left ( \frac{ \mathbb{e}^4 W^2}{4 U \Delta^2} \right )
\right \}.
\label{expans}
\end{align}
where $\mathbb{e}$ is the Napier's constant and  ${\cal W}(z)$ is the  Lambert ${\cal W}$-function 
(or so called ProductLog function) which obeys $ {\cal W}(z) \mathbb{e}^{{\cal W}(z)}=z$.~\cite{wolfram}

The above set of equations (\ref{newapp1})-(\ref{newapp4})
can be employed also to investigate the
critical value $U=U_c$ at which the bandgap $\Delta_{\rm G}$ closes,
i.e. $\Delta_{\rm G}(U_c)=0$. Note that, by definition,
$\Delta_{\rm G}=\omega_{\rm c}^*$ 
which, as discussed above, does {\em not} imply $\widetilde {\omega}_{\rm c}^*=0$.
We have thus $\alpha^*_c(U_{\rm c})=1/U_{\rm c}$,
$\Delta_{\rm c}(U_{\rm c})=\Delta/2$ and
\begin{align}
\alpha^*_{\rm c}(U_{\rm c})
&=
\ln\left[
1+\frac{4W^2}{\Delta^2}
\right].
\end{align}
As a consequence, we can get the final analytical result:
\begin{align}
U_{\rm c}
&=
\frac{1}
{
\ln\left[
1+\frac{\displaystyle 4W^2}{\displaystyle \Delta^2}
\right]
}.
\label{c2}
\end{align}

Alternatively, we can look back at 
Eqs. (\ref{e:ap2})-(\ref{d:ap2}) in the complex space,
and consider the case $\omega=0$.
For symmetry we have that
$\Sigma_I(\omega=0)=-i\Gamma$
is a purely imaginary quantity, whereas
$\Sigma_z(\omega=0)=\sigma$
is a purely real one.
As a byproduct, also the quantity
$\alpha(\omega=0)$
appears to be a purely real one.
We can write a self-consistent set of equations:
\begin{align}
\Gamma
&=
\Gamma
U
\ln\left[
1+\frac{W^2}{(\Delta+\sigma)^2+\Gamma^2}
\right],
\\
\sigma
&=
-
(\Delta+\sigma)
U
\ln\left[
1+\frac{W^2}{ (\Delta+\sigma)^2+\Gamma^2}
\right],
\end{align}
For small $U$ the only possible solution is $\Gamma=0$.
A finite $\Gamma$ is sustained at a critical value $U_{\rm c}$
which can be found by the conditions:
\begin{align}
1
&=
U_{\rm c}
\ln\left[
1+\frac{W^2}{(\Delta+\sigma)^2}
\right],
\\
\sigma
&=
-
(\Delta+\sigma)
U_{\rm c}
\ln\left[
1+\frac{W^2}{(\Delta+\sigma)^2}
\right].
\end{align}
This set of equations imply $\sigma=-\Delta/2$, and an analytical expression for $U_{\rm c}$
\begin{align}
1
&=
U_{\rm c}
\ln\left[
1+\frac{4W^2}{\Delta^2}
\right],
\end{align}
which corresponds to Eq. (\ref{c2}).

\end{document}